\documentclass[twocolumn,showpacs,preprintnumbers,amsmath,amssymb]{revtex4}

\usepackage{graphicx}% Include figure files
\usepackage{dcolumn}% Align table columns on decimal point
\usepackage{bm}% bold math

\begin{document}
\title{Modeling Human Dynamics with Adaptive Interest}

\author{Xiao-Pu Han$^{1}$}
%\email{hxp@mail.ustc.edu.cn}
\author{Tao Zhou$^{1,2}$}
\email{zhutou@ustc.edu}
\author{Bing-Hong Wang$^{1,3}$}
%\email{bhwang@ustc.edu.cn}

\affiliation{$^{1}$ Department of Modern Physics, University of
Science and Technology of China, Hefei 230026 P. R. China\\ $^{2}$
Department of Physics, University of Fribourg, CH-1700, Fribourg,
Switzerland\\ $^{3}$ Shanghai Institute for Systemic Sciences,
Shanghai, 200093, P. R. China }

\date{\today}

\begin{abstract}
Recently, increasing empirical evidence indicates the extensive
existence of heavy tails in the interevent time distributions of
various human behaviors. Based on the queuing theory, the Barab\'asi
model and its variations suggest the highest-priority-first protocol
a potential origin of those heavy tails. However, some human
activity patterns, also displaying the heavy-tailed temporal
statistics, could not be explained by a task-based mechanism. In
this paper, different from the mainstream, we propose an
interest-based model. Both the simulation and analysis indicate a
power-law interevent time distribution with exponent -1, which is in
accordance with some empirical observations in human-initiated
systems.
\end{abstract}


\maketitle

\section{INTRODUCTION}
Human behavior, as an academic issue in science, has a history of
about one century from Watson \cite{Watson1914}. As a joint interest
of sociology, psychology and economics, human behavior has been
extensively investigated during the last decades. However, due to
the complexity and diversity of our behaviors, the in-depth
understanding of human activities is still a long-standing challenge
thus far. Actually, in most of the previous works, the individual
activity pattern is usually simplified as a completely random
point-process, which can be well described by the Poisson process,
leading to an exponential interevent time distribution \cite{Hai1}.
That is to say, the time difference between two consecutive events
should be almost uniform, and the long gap is hardly to be observed.
However, recently, the empirical studies on e-mail \cite{Bar2} and
surface mail \cite{Oliv3} communication show a far different
scenario: those communication patterns follow non-Poisson
statistics, characterized by bursts of rapidly occurring events
separated by long gaps. Correspondingly, the interevent time
distribution has a much heavier tail than the one predicted by an
exponential distribution. The heavy tails have also been observed in
many other human behaviors \cite{Goh2008,Zhou2008}, including market
transaction \cite{Ple4,Mas5}, web browsing \cite{Dez6}, movie
watching \cite{Zhou7}, short message sending \cite{Hong8}, and so
on. The increasing evidence of non-Poisson statistics of human
activity pattern highlights a question: what is the origin of those
heavy tails? Based on the queuing theory, Barab\'asi \emph{et al.}
proposed a simple model \cite{Bar2,Vqz9,Vqz10} where the individual
executes the highest-priority task first, and they suggested the
highest-priority-first (HPF) protocol a potential origin of those
heavy tails.

The queuing model gets a great success in explaining the heavy tails
in many human-oriented dynamics. However, some other human activity
patterns, also displaying the similar heavy-tailed phenomenon, could
not be explained by a task-based mechanism. For example, the actions
on browsing webs \cite{Dez6}, watching on-line movies \cite{Zhou7},
and playing on-line games \cite{Hen11} are mainly driven by the
personal interests, which could not be treated as tasks needing to
be executed. The in-depth understanding of the non-Poisson
statistics in those interest-driven systems requires a new model out
of the perspective of queuing theory. In this paper, different from
the mainstream task-based models, we propose an interest-based
model. Both the simulation and analysis indicate a power-law
interevent time distribution with exponent -1, which is in
accordance with some empirical human-initiated systems.

\section{MODEL}
Before introducing the mathematical rules of our model, let us think
of the changing process of our interests on web browsing according
to our daily experiences. If a person has a long period not browsing
the web, an accidental visit may give him a good feeling and wake
his interest on the web browsing. Next, during the actions, the good
feeling continues and the frequency of web browsing may increase.
Then, if the frequency is too high, he may worry about it, thus
reduces those browsing actions. Such similar experiences can be
found in many other daily actions, such as playing games, seeing
movies, and so on. In a word, we usually adjust the frequency of the
daily actions according to our interest: greater interest will lead
to higher frequency, and vice versa. Some simple assumptions
extracted from our daily experiences are as follows: Firstly, for a
given interest-driven behavior, each action will change the current
interest, while the frequency of actions depends on the interest. It
likes an active walker \cite{Lam2005,Lam2006}, whose motion is
affected by the energy landscape, while the motion track could
simultaneously change the landscape. Secondly, the interevent time
$\tau$ has two thresholds: when $\tau$ is too small (i.e., events
happen too frequently), the interest will be depressed, thus the
interevent time will increase; while if the time gap is too long, we
will increase the interest to mimic its resuscitation induced by a
casual action.

According to those assumptions, we propose an interest-based model
as follows: (i) The time is discrete and labelled by $t = 0, 1, 2,
\cdots$, the occurring probability of an event at time step $t$ is
denoted by $r(t)$. The time interval between two consecutive events
is call the interevent time and denoted by $\tau$. (ii) If the
$(i+1)$th event occurred at time step $t$, the value of $r$ is
updated as $r(t+1) = a(t)r(t)$, where
\begin{equation}
    a(t)=\left\{
    \begin{array}{cc}
a_{0}, & \tau _{i} \leq  T_{1}, \\
a_{0}^{-1}, & \tau _{i}\geq  T_{2}, \\
a(t - 1),& T_{1} < \tau _{i} < T_{2}.
    \end{array}
    \right.
\end{equation}
If no event occurred at time step $t$, we set $a(t)=a(t-1)$, namely
$a(t)$ keeps unchanged. In this definition, $T_{1}$ and $T_{2}$ are
two thresholds satisfied $T_1 \ll T_2$, $\tau_{i}$ is the time
interval between the $(i+1)$th and the $i$th events, and $a_{0}$ is
a parameter controlling the changing rate of occurrence probability
($0<a_{0}<1$). If no event happens, $r$ will not change. Clearly,
simultaneously enlarge (by the same multiple) $T_1$, $T_2$, and the
minimal perceptible time, the statistics of this system will not
change. Therefore, without lose of generality, we set $T_1=1$.

\begin{figure}
  \includegraphics[width=8.6cm]{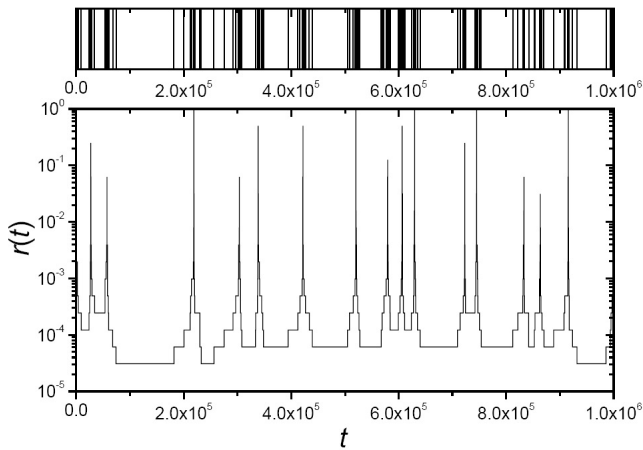}\\
  \caption{
 (upper panel) The succession of events predicted by the present
 model. The total number of events shown here is 375 during $10^{6}$ time
 steps.
  (lower panel) The corresponding changes of $r(t)$. The data points are obtained
  with the parameters $a_{0} = 0.5$ and $T_{2} = 10^{4}$.}
\end{figure}

\section{SIMULATION AND ANALYSIS}

In the simulations, the initial value of $r$ is set as
$r_0=r(t=0)=1.0$, which is also the possibly maximal value of $r(t)$
in the whole simulation process. As shown in Fig. 1, the succession
of events predicted by the present model exhibits very long inactive
periods that separate the bursts of rapidly occurring events, and
the corresponding $r(t)$ shows a clearly seasonal property
(quasi-periodic behavior). Actually, in a period, the maximal and
minimal values of $r(t)$ respectively determined by $T_1$ and $T_2$
as $r_{\texttt{max}} \sim T_1^{-1}$ and $r_{\texttt{min}} \sim
T_2^{-1}$. This quasi-periodic property will be applied in the
further analysis. Note that, in a specific quasi-period,
$r_{\texttt{max}}$ can be smaller than $T_1^{-1}$ and
$r_{\texttt{min}}$ can be smaller than $T_2^{-1}$. It is because
$\tau\leq T_1$ could happen when $r(t)<T_1^{-1}$ and $\tau\leq T_2$
could happen when $r(t)\leq T_2^{-1}$.

Figure 2 reports the simulation results with tunable $T_2$ and
$a_0$. Given $a_0=0.5$, if $T_{2} \gg T_{1}$, the interevent time
distribution generated by the present model displays a clearly power
law with exponent -1; while if $T_2$ is not sufficiently large, the
distribution $P(\tau)$ exhibits a departure from a power-law form
with a cut-off in its tail. Correspondingly, given sufficiently
large $T_2$, the effect of $a_0$ is very slight, thus can be
ignored.

\begin{figure}
  \includegraphics[width=8.6cm]{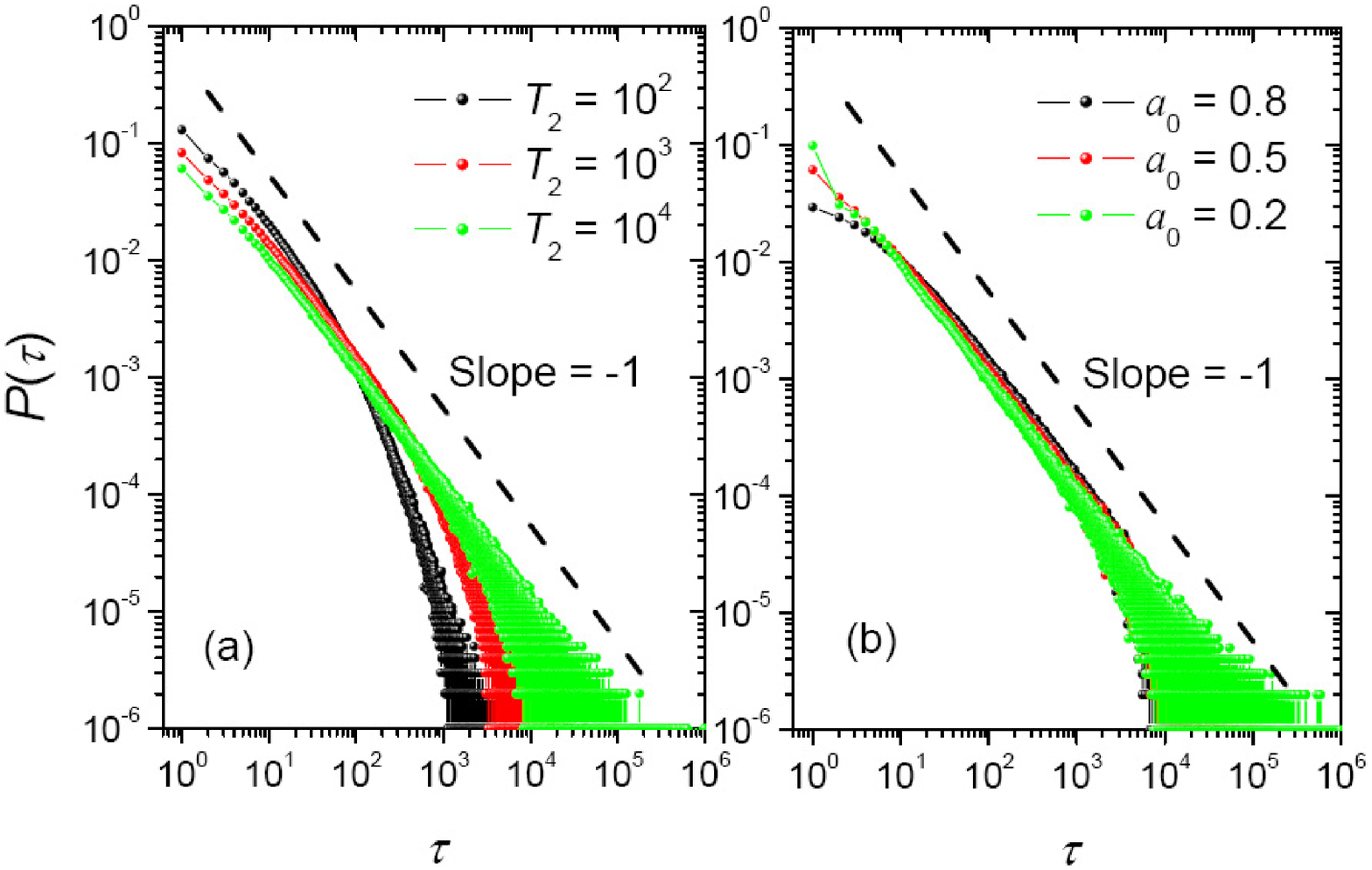}\\
  \caption{(Color online) The interevent time distributions in log-log plots. (a) Given $a_{0}=0.5$,
  $P(\tau)$ for different $T_{2}$, where the black, red and green curves denote the cases of $T_{2} = 10^{2}$, $10^{3}$ and
$10^{4}$, respectively. (b) Given $T_{2} = 10^{4}$, $P(\tau)$ for
different $a_{0}$, where the black, red and green curves denote the
cases of $a_{0} = 0.8$, $0.5$ and $0.2$, respectively. The black
dash lines in both (a) and (b) have slope -1. Each distribution
contains $10^{6}$ events. }
\end{figure}

Taking into account the quasi-periodic property of $r(t)$, we raise
two approximated assumptions before analytical derivation: (i) The
statistical property of $P(\tau)$ is the same as that in a single
period; (ii) Within one period, the statistical property of
$P(\tau)$ in the $r$-increasing half is the same as that in the
$r$-decreasing half. In the reducing process, $r(t) =
r_{m}a_{0}^{i}$, where $i = 0, 1, 2, \cdot\cdot\cdot, I$. The
integer $I$ denotes the number of the events in the reducing process
(also the number of different values of $r(t)$), whose value is
about
\begin{equation}
I\approx -\log_{a_{0}}(T_{2}/T_{1})
\end{equation}
since $r_{\texttt{max}} \sim T_1^{-1}$ and $r_{\texttt{min}} \sim
T_2^{-1}$. $r_{m}$ is the initial value (it is also the maximum
value) of $r(t)$ in a reducing process. Note that, for different
reducing processes, the values of $r_{m}$ are not always the same.
Though $r_m$ has the same order of magnitude with $T_{1}^{-1}=1.0$,
its value can be less than $T_{1}^{-1}$ in a specific process. The
average value of $r_{m}$ will be calculated later in this paper.

If the current occurring probability is $r(t)=r_ma_0^i$, the
probability that the next event will happen at the time $t+\tau$ is:
\begin{equation}
Q(\tau)=(1 - r_{m}a_{0}^{i})^{\tau - 1} r_{m}a_{0}^{i}.
\end{equation}
Considering every value of $r(t)$ in the reducing process, the
interevent time distribution of the reducing process is:
\begin{equation}
P(\tau)=I^{-1}\sum_{i = 0}^{I}(1 - r_{m}a_{0}^{i})^{\tau - 1}
r_{m}a_{0}^{i}.
\end{equation}
According to the approximated assumptions above, the interevent time
distribution of all the successions can also be expressed by Eq.
(4), which can be approximately rewritten in a continuous form, as:
\begin{equation}
P(\tau) \approx I^{-1}\int_{0}^{I}(1 - r_{m}a_{0}^{x})^{\tau - 1}
r_{m}a_{0}^{x}dx.
\end{equation}
Therefore, $P(\tau)$ can be further expressed as:
\begin{equation} P(\tau) \approx - [(1 - r_{m}a_{0}^{I})^\tau - (1 -
r_{m})^\tau](\ln a_{0})^{-1}I^{-1}\tau^{-1}.
\end{equation}
From Eq. (6), for a fixed $r_{m}$, when $I$ is large enough (it is
equivalent to the condition $T_{2}\gg T_1$), $P(\tau)$ has a
power-law tail with exponent -1. In addition, this analytical result
also provides an explanation about the departure from a power law
when $T_2$ is not sufficiently large.

\begin{figure}
  \includegraphics[width=8.6cm]{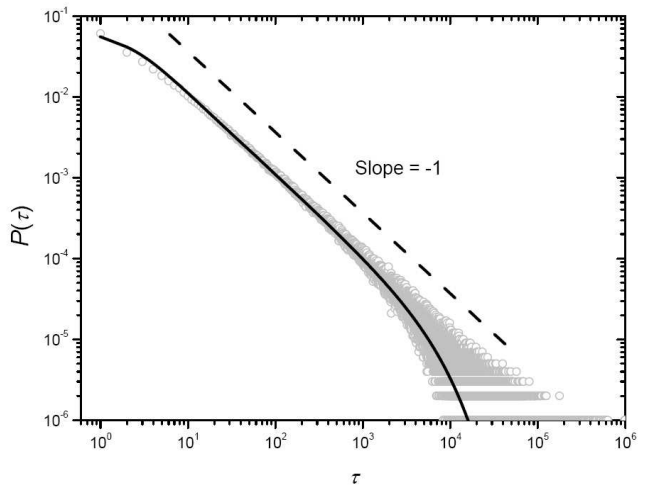}\\
  \caption{The comparison of the analytical (black solid line) and numerical (gray circles) results of interevent time distribution. The numerical
  data are obtained with parameters $r_{0} = 1.0$, $a_{0} = 0.5$ and $T_{2} = 10^{4}$. The analytical
  results are calculated by Eq. (6) with $a_{0} = 0.5$, $I = 13$ and $r_{m} = 0.50$.
  The black dash line has slope -1. The numerical results contain $10^{6}$ events. }
\end{figure}

As discussed before, for different reducing processes of $r(t)$, the
possible values of $r_{m}$ are not always the same (see also the
lower panel of Fig. 1: for different quasi-periods, the maximum
values of $r(t)$ are different). Since the order of magnitude of
$r_{m}$ is comparable with $T_{1}^{-1}=1.0$ (it is equal to
$r_{0}$), the minimum value of $r(t)$, $r_{m}a_{0}^{I}$, has the
same order of magnitude with $r_{0}a_{0}^{I}$. Making the
approximated assumption that the minimum value of $r(t)$ is given by
$r_{0}a_{0}^{I}$ in a $r$-increasing process, and the maximum value
of $r(t)$ in the next $r$-decreasing process is $r_0a_0^k$
($r_0a_0^k$ is also the start point in the next decreasing process),
then the probability density of $k$ reads
\begin{equation}
\Omega(k) = r_{0}a_{0}^{k}\prod_{i = 0}^{I - k -1}(1 - r_{0}a_{0}^{I
- i}).
\end{equation}
Therefore, the average value of $r_{m}$ is
\begin{equation}
\langle r_{m}\rangle = \sum_{k = 0}^{I - 1}r_{0}a_{0}^{k}\Omega(k) =
\sum_{k = 0}^{I - 1}(r_{0}a_{0}^{k})^{2}\prod_{i = 0}^{I - k -1}(1 -
r_{0}a_{0}^{I - i}).
\end{equation}
This average value of $r_{m}$ calculated by Eq. (8), as well as the
integer part of $-\log_{a_{0}}(T_{2}/T_{1})$ (as the approximation
of $I$), can be directly used in the approximate calculations of Eq.
(6). Given $r_{0} = 1.0$, $a_{0} = 0.5$, $T_{2} = 10^{4}$ and $T_{1}
= 1$, one obtains $I\approx-\log_{a_{0}}(T_{2}/T_{1})=13$, and
$\langle r_m\rangle \approx 0.50$ by Eq. (8). Accordingly, Fig. 3
reports the comparison of analytical and simulation results, which
are well in accordance with each other.

\section{CONCLUSION AND DISCUSSION}
A novel model on human dynamics is proposed in this paper. Different
from the mainstream queuing models, the current model is driven by
the personal interests. In this model, the frequency of events are
determined by the interest, while the interest are simultaneously
affected by the occurrence of events. This interplay working
mechanism, similar to the active walk \cite{Lam2005,Lam2006}, is a
genetic origin of complexity of many real-life systems. The rules in
the current model are extracted from our daily life, and both the
analytical and simulation results agree well with the empirical
observation, such as the activity pattern of web browsing
\cite{Dez6}. Our work indicates a simple and universal mechanism in
human dynamics, that is, a people could adaptively adjust their
interest on a specific behavior (e.g. watching TV, browsing web,
playing on-line game, etc.), which leads to a quasi-periodic change
of interest, and this quasi-periodic property eventually gives raise
to the departure of Poisson statistics.

Besides the HPF protocol and the current model, there are also some
other mechanisms that can lead to a power-law interevent time
distribution. For example, Hidalgo \cite{Hidalgo2006} pointed out
that a Poissonian individual with characteristic time varying
randomly in time could generate a power-law interevent time
distribution with exponent -2. In addition, V\'azquez
\cite{Vazquez2007} showed that if the current executing rate is
linearly correlated with the average executing rate in a immediate
predecessor period, the interevent time distribution will follow a
power-law form.

Note that, although in the recent empirical works, the power-law
form is widely used to fit the interevent time distribution of human
behaviors, there exists a debate about the choice of fitting
functions for this distribution in the e-mail communication
\cite{Stouffer2005,Barabasi2005}. Actually, a candidate, namely
\emph{log-normal distribution}, has also been suggested
\cite{Stouffer2005} to describe the non-Poisson temporal statistics
of human activities. The \emph{stretched exponential distribution}
\cite{Laherrere1998,Zhang2006}, interpolating between a power law
and an exponential form, serves as another candidate (see, for
example, the distribution of interevent time between two consecutive
transactions initiated by a stock broker \cite{Vqz10}). A clear
understanding of the tails in the interevent time distribution asks
for in-depth exploration on empirical data in the future.

The concept and methodologies related to the statistics of the
interevent time can also find its applications in some other
systems. For example, similar statistical analysis can be addressed
on the spacing between the consecutive occurrences of the same
letter in written text \cite{Goh2008}, and the time difference
between successive events above a certain threshold (i.e., extreme
events) \cite{Bogachev2007}.

Finally, we point out some limitations in the current model.
Firstly, it can only generate the power-law interevent time
distribution with exponent -1, which does not agree with some real
human-initiated systems with different power-law exponents.
Secondly, we assume that the changing rate of the occurring
probability, $a_0$, is fixed as a constant in every rising or
decaying process. This assumption is very ideal, and we could not
find any support from the empirical data. Third, as stated by
Kentsis \cite{Kentsis2006}, there are countless ingredients
affecting the human dynamics, and for most of them, we do not know
their impacts. Those ingredients, such as the social content, the
semantic content and the periodicity due to circadian and weekly
cycles, have not been considered in the present model, neither the
HPF protocol. However, although this model is rough and may contain
some artificial assumptions, it provides a start point of modeling
interest-based human dynamics. The human-initiated systems are the
most complex systems, and there must be many underlying mechanisms
having not been discovered yet. We believe our model could highlight
the readers in this rapidly growing area.

\begin{acknowledgments}
We acknowledge the the useful discussion with Wei Hong and
Shuang-Xing Dai, this work is partially supported by the National
Natural Science Foundation of China (Grant Nos. 10472116 and
10635040), and the 973 Program 2006CB705500.
\end{acknowledgments}


\begin{thebibliography}{ref1}
\bibitem{Watson1914} J. B. Watson, Psychological Review {\bf 20}, 158 (1913).
\bibitem{Hai1} F. A. Haight, \emph{Handbook of the Poisson Distribution} (Wiley, New York, 1967).
\bibitem{Bar2} A. -L. Barab\'{a}si, Nature (London) \textbf{435}, 207 (2005).
\bibitem{Oliv3} J. G. Oliveira, and A. -L. Barab\'{a}si, Nature (London) \textbf{437}, 1251 (2005).
\bibitem{Goh2008} K. -I. Goh, and A. -L. Barab\'{a}si, EPL {\bf 81}, 48002 (2008).
\bibitem{Zhou2008} T. Zhou, X. -P. Han, and B. -H. Wang, arXiv: 0801.1389.
\bibitem{Ple4} V. Plerou, P. Gopikrishnan, L. A. N. Amaral, X. Gabaix, and H. E. Stanley, Phys. Rev. E \textbf{62}, 3023(R)(2000).
\bibitem{Mas5} J. Masoliver, M. Montero, and G. H. Weiss, Phys. Rev. E \textbf{67}, 021112 (2003).
\bibitem{Dez6} Z. Dezs\"{o}, E. Almaas1, A. Luk\'{a}cs, B. R\'{a}cz, I. Szakad\'{a}t, and A. -L. Barab\'{a}si, Phys. Rev. E \textbf{73}, 066132 (2006).
\bibitem{Zhou7} T. Zhou, H. A. -T. Kiet, B.J. Kim, B. -H. Wang, and P. Holme, EPL {\bf 82}, 28002 (2008).
\bibitem{Hong8} W. Hong, X.- P. Han, T. Zhou, and B. -H. Wang, arXiv: 0802.2577.
\bibitem{Vqz9} A. V\'{a}zquez, Phys. Rev. Lett. \textbf{95}, 248701 (2005).
\bibitem{Vqz10} A. V\'{a}zquez,  J. G. Oliveira,  Z. Dezs\"{o}, K. -I. Goh, I. Kondor, and A. -L. Barab\'{a}si, Phys. Rev. E \textbf{73}, 036127 (2006).
\bibitem{Hen11} T. Henderson, and S. Nhatti, Proc. 9th ACM Int. Conf. on Multimetia, pp. 212 (ACM Press, 2001).
\bibitem{Lam2005} L. Lam, Int. J. Bifurcation \& Chaos {\bf 15}, 2317 (2005).
\bibitem{Lam2006} L. Lam, Int. J. Bifurcation \& Chaos {\bf 16}, 239 (2006).
\bibitem{Hidalgo2006} C. Hidalgo, Physica A {\bf 369}, 877 (2006).
\bibitem{Vazquez2007} A. V\'{a}zquez, Physica A {\bf 373}, 747 (2007).
\bibitem{Stouffer2005} D. B. Stouffer, R. D. Malmgren, and L. A. N. Amaral, arXiv: physics/0510216.
\bibitem{Barabasi2005} A. -L. Barab\'{a}si, K. -I. Goh, and A. V\'azquez, arXiv: physics/0511186.
\bibitem{Laherrere1998} J. Laherrere and D. Sornette, Eur. Phys. J. B {\bf 2}, 525 (1998).
\bibitem{Zhang2006} P. P. Zhang, K. Chen, Y. He, T. Zhou, B. B. Su, Y. D. Jin, H. Chang, Y. P. Zhou, L. C. Sun, B. H. Wang, D. R. He, Physica A {\bf 360}, 599 (2006).
\bibitem{Bogachev2007} M. I. Bogachev, J. F. Eichner, and A. Bunde, Phys. Rev. Lett. {\bf 99}, 240601 (2007).
\bibitem{Kentsis2006} A. Kentsis, Nature (London) {\bf 441}, E5 (2006).


\end{thebibliography}
\end{document}